\documentclass[twocolumn,showpacs,preprintnumbers,amsmath,amssymb,floatfix]{revtex4}
\usepackage{graphicx}% Include figure files
\usepackage{dcolumn}% Align table columns on decimal point
\usepackage{bm}% bold math

\newcommand{\pom}{\tt I\! P}
\newcommand{\beq}{\begin{equation}}
\newcommand{\eeq}{\end{equation}}

\begin{document}

\title{Diffractive Hadroproduction of Dijets and W's at the Tevatron Collider\\
and the Pomeron Structure Function }

\author{R. J. M. Covolan}
\author{M. S. Soares}
\altaffiliation{Presently at DESY, Hamburg, Germany.}
\affiliation{Instituto de F\'{\i}sica {\em Gleb Wataghin} \\
{\small Universidade Estadual de Campinas, Unicamp} \\
{\small 13083-970 \ Campinas \ SP \ Brazil }}

\begin{abstract}

Results from a phenomenological analysis of dijet and $W$ hard diffractive
hadroproduction at Fermilab Tevatron energies are reported. The theoretical
framework employed here is a modified version of the Ingelman-Schlein
approach which includes DGLAP-evolved structure functions. Different from
what has been achieved by the DESY $ep$ HERA reactions, a reasonable overall
description of such diffractive hadron processes is obtained only when a 
complex,
quark-rich Pomeron structure function is employed in the calculation.

\end{abstract}

%\vspace{1cm}


\maketitle

%\newpage

\section{Introduction}

Regge phenomenology is well known for providing a suitable and economical
theoretical framework for the description of {\it soft} hadron diffractive
processes at high energies. Single Pomeron exchange plus a few secondary Reggeon
contributions are enough to describe a variety of hadronic reactions (see, for
instance, \cite{covolan} and references therein).

The situation becomes much more intricate when one considers {\it hard} diffractive
processes by which, according to the Ingelman-Schlein (IS) picture \cite{ingelman},
the Pomeron structure itself is probed. Difficulties arise when one
tries to obtain a unified description for diffractive processes starting with
both electron- or positron-proton ($ep$) and antiproton-proton (${\bar p}p$) 
collisions,
respectively, studied at the DESY $ep$ HERA and Tevatron colliders.
Although several theoretical approaches have successfuly been employed to describe
different aspects of hard diffraction revealed by the $ep$ HERA reactions
\cite{halina}, some of them based on Regge theory, diffractive
hadroproduction continues to be one of the most challenging topics in hadron
dynamics.

In effect, most of the theoretical approaches which are able to describe the HERA 
data are not readily translatable to diffractive hadron physics. Models based on the
Regge theory, in 
particular, are presumably affected by a lack of validity for QCD
factorization in the hadronic diffraction domain \cite{collins}. In spite of these
difficulties, such models establish the phenomenological picture most of the 
event generators currently employed in data analysis of hard diffractive 
hadroproduction are based upon. Probably this is so because these models have  
been able to provide an effective description for such processes. In fact, that
is one of the underlying assumptions of the present paper.

This Brief Report is a sequel to a previous work \cite{mara} in which we have
tried to perform a global analysis by a modified version of the IS model,
including processes initiated by both $ep$ and ${\bar p}p$ collisions.
By that time, the available data were not so stringent such that one could
speculate about an unique model to be sufficient. Since then, more and more
precise data of diffractive deep inelastic scattering (DIS) have imposed severe
restrictions on the Pomeron structure function making more evident the
impossibility to readily transfer the partonic densities so obtained to
hadronic process calculations.

Our group's analysis of the diffractive DIS data have
shown that, at low values of the QCD evolution scale, the Pomeron is predominantly
composed by gluons with a hard distribution, in agreement with other studies
(see \cite{batista} and references therein). This result is corroborated by the
calculations of diffractive cross sections for photoproduction \cite{foto} and
electroproduction \cite{eletro} of dijets, which yield good agreement with the data
once a hard gluonic Pomeron is assumed.

The advent of new data produced by
the D0 Collaboration \cite{dzero} have motivated us to perform a new analysis, this
time restricted to the Tevatron data.

Thus, we report here results of a study on diffractive hadroproduction of dijets
and W's by using the IS model, with Dokhshitzer-Gribov-Lipatov-Altarelli-Parisi
(DGLAP) evolution \cite{dglap} included, but disconnected
from the HERA data analysis just mentioned \cite{batista}. In fact, in spite of our
efforts it was impossible to reconcile both analyses. We take this failure to
produce a unified description as an additional indication of the theoretical
problems alluded before.

The interesting point, however, is that if one takes the Pomeron as predominantly
composed by quarks at a low QCD evolution scale, a reasonable overall description
of diffractive hadroproduction data is achieved. This is what is shown below.

%
%%%%%%%%%%%%%%%%%%%%%%%%%%%%%%%%%%%%%%%%%%%%%%%%%%%%%%%%%%%%%%%%%%%%%%%%%%
%

\section{Cross Sections}

Our starting point is the generic cross section for a process in which partons
of two hadrons, $A$ and $B$, interact to produce jets (or $W$s),
$
A + B \rightarrow \ Jets\ (W) + X,
$
that is
\begin{eqnarray}
\nonumber
d \sigma =&& \sum_{a,b,c,d} f_{a/A}(x_a,
\mu^2)\ dx_a\ f_{b/B}(x_b, \mu^2)\ dx_b \\ 
&&\times\ \frac{d\hat{\sigma}_{ab\rightarrow cd (W)}}{d\hat{t}}\ d\hat{t}.
\label{gen}
\end{eqnarray}
From this very basic expression we derive all of the others necessary to
describe the specific processes we are interested in.

\subsection{Diffractive dijet production from a single Pomeron exchange}

In the case of dijet production, the cross section
can be put in terms of the
dijet rapidities ($\eta, \eta'$) and transversal energy $E_T$:
\begin{eqnarray}
\nonumber
\left(\frac{d\sigma}{d\eta}\right)_{jj}=&&\sum_{\rm partons}
\int_{E_{T\rm{min}}}^{E_{T\rm{max}}}
d E_T^2 \int_{\eta'_{\rm{min}}}^{\eta'_{\rm{max}}} d\eta' \\ 
&&\times\ x_a f_{a/A}(x_a,\mu^2)\ x_b  f_{b/B}(x_b,\mu^2)
\left(\frac{d\hat{\sigma}}{d\hat{t}}\right)_{jj}
\label{modpar}
\end{eqnarray}
where
\begin{equation} 
x_a = \frac{E_T}{\sqrt{s}}(e^{-\eta}+e^{-\eta^{\prime}}), \ \ \ \ \ \ \ x_b
= \frac{E_T}{\sqrt{s}}(e^{\eta}+e^{\eta^{\prime}}),  
\label{xbj}
\end{equation}
with
\begin{eqnarray}
\ln{\frac{E_T}{\sqrt{s}-E_T\ e^{-\eta }}} \leq \eta ' \leq
\ln{\frac{\sqrt{s}-E_T\ e^{-\eta }}{E_T}} 
\end{eqnarray}
and
\begin{eqnarray}
E_{T\ \rm{max}}=\frac{\sqrt{s}}{e^{-\eta }+e^{\eta }},
\label{etmax}
\end{eqnarray}
being that $E_{T\ \rm{min}}$ and the $\eta$ range are determined by
the experimental cuts.

Equations.~(\ref{modpar})-(\ref{etmax}) express the usual leading-order QCD procedure 
to
obtain the {\em non-diffractive} dijet cross section (next-to-leading-order
contributions are not essential for the present purposes; see Ref.~\cite{mara}).
In order to obtain the corresponding expression for
{\em diffractive} processes, we assume that one of the hadrons, say
hadron $A$, emits a Pomeron whose partons interact with partons of the hadron $B$.
Thus the parton distribution  $x_a f_{a/A}(x_a, \mu^2)$ in
Eq.~(\ref{modpar}) is replaced by the convolution between a putative
distribution of partons in the Pomeron, $\beta f_{a/{\tt I\! P}}(\beta,
\mu^2)$, and the ``emission rate" of Pomerons by $A$, $f_{{\tt I\!
P}}(x_{{\tt I\! P}},t)$. This last quantity, $f_{{\tt I\! P}}(x_{{\tt I\!
P}},t)$, is the so-called Pomeron flux factor whose explicit formulation in
terms of Regge theory is given ahead. The whole procedure implies that
\begin{eqnarray}
\nonumber
\label{convol}
x_a f_{a/A}(x_a, \mu^2) =&& \int dx_{{\tt I\! P}} \int d\beta \int dt\
f_{{\tt I\! P}}(x_{{\tt I\! P}},t) \\
&& \times\ \beta\ f_{a/{\tt I\! P}}(\beta, \mu^2)\
\delta(\beta-\frac{x_a}{x_{{\tt I\! P}}}),
\end{eqnarray}
and, defining $g (x_{{\tt I\! P}}) \equiv \int_{-\infty}^0 dt\
f_{{\tt I\! P}}(x_{{\tt I\! P}},t)$, one obtains
\begin{eqnarray}
\label{convoP}
x_a f_{a/A}(x_a, \mu^2)\ =\ \int dx_{{\tt I\! P}} \
g(x_{{\tt I\! P}})\ {\frac{x_a}{x_{{\tt I\! P}}}} f_{a/{\tt I\!P}}
({\frac{x_a}{x_{{\tt I\! P}}}}, \mu^2).
\end{eqnarray}
By inserting the above structure function into Eq.~(\ref{modpar}) one obtains
the cross section for diffractive hadroproduction of dijets via a single
Pomeron exchange as
\begin{widetext}
\begin{eqnarray}
\left(\frac{d\sigma_{SPE}}{d\eta}\right)_{jj}=\sum_{a,b,c,d}
\int_{E_{T_{min}}}^{E_{T_{max}}} dE_T^2
\int_{\eta'_{min}}^{\eta'_{max}} d\eta' 
\int_{x_{{\tt I\! P}_{min}}}^{x_{{\tt I\! P}_{max}}}
dx_{\tt I\! P}\ g(x_{\tt I\! P})\  \beta_a f_{a/{\tt I\! P}}(\beta_a, \mu^2)
\ x_b f_{b/\bar{p}}(x_b, \mu^2)\
\left(\frac{d\hat{\sigma}_{ab\rightarrow cd}}{d\hat{t}}\right)_{jj},
\label{dsigjato}
\end{eqnarray}
\end{widetext}
where $\beta_a = {x_a}/x_{\tt I\! P}$ with $x_a$ and $x_b$ given by 
Eq.~(\ref{xbj}), and $x_{{\tt I\! P}_{min}}$ and $x_{{\tt I\! P}_{max}}$
established by the experimental cuts.

%
%%%%%%%%%%%%%
%

\subsection{Diffractive Hadroproduction of $W^{\pm}$}

$W^{\pm}$ diffractive production is here considered by the reaction
$p + {\bar p} \rightarrow p + \ W (\rightarrow e\ \nu ) + \ X.$
It is assumed that a Pomeron emitted by a proton in
the positive $z$ direction interacts with a $\bar p$ producing $W^{\pm}$
that subsequently decays into $e^{\pm}\ \nu$. The detection of this reaction is
triggered by the lepton ($e^+$ or
$e^-$) that appears boosted towards negative $\eta$ (rapidity) in coincidence
with a rapidity gap in the right hemisphere.

By using the same concept of the convoluted structure function, the
diffractive cross section for the inclusive lepton production for 
this process becomes \cite{nota1}
\begin{widetext}
\begin{eqnarray}
\left(\frac{d\sigma_{SD}}{d\eta_e}\right)_{lepton}= \sum_{a,b}
\int {\frac{dx_{{\tt I\! P}}}{x_{{\tt I\! P}}}}\ g(x_{{\tt I\!P} })
\int dE_T \ f_{a/{\tt I\! P}}(x_a, \mu^2)\ f_{b/\bar{p}}(x_b, \mu^2)\
\left[\frac{ V_{ab}^2\ G_F^2}{6\ s\ \Gamma_W}\right]\ \frac{\hat{t}^2}
{\sqrt{A^2-1}}
\label{dsw}
\end{eqnarray}
\end{widetext}
where
\begin{equation}
x_a = \frac{M_W\ e^{\eta_e}}{(\sqrt{s}\ x_{{\tt I\! P}})}\ \left[A \pm
\sqrt{(A^2-1)}\right],
\label{xaw}
\end{equation}
\begin{equation}
x_b = \frac{M_W\ e^{-\eta_e}}{\sqrt{s}}\ \left[A \mp \sqrt{(A^2-1)}\right],
\label{xbw}
\end{equation}
and 
\begin{equation}
\hat{t}=-E_T\ M_W\ \left[A+\sqrt{(A^2-1)}\right]
\label{tw}
\end{equation}
with $A={M_W}/{2 E_T}$. The upper signs in Eqs.~(\ref{xaw}) and (\ref{xbw})
refer to $W^+$ production (that is, $e^+$ detection). The corresponding
cross section for $W^-$ is obtained by using the lower signs and ${\hat t}
\leftrightarrow {\hat u}$ (see the Appendix in \cite{mara}).

\subsection{The Pomeron Flux Factor}

An important element of this approach is the Pomeron flux factor, introduced in
Eq.~(\ref{convol}). It has some peculiar aspects that deserve to be pointed out.

First of all, the expression for this term was originally proposed to be taken from
the invariant cross section of (soft) diffractive dissociation processes as it is
given by the Triple Pomeron model \cite{ingelman}. The rationale for that 
can be put in terms of an analogy with the photon flux factor, this one derived 
from QED.
The basic idea is that, similarly to what happens to the electron (or positron) in
photoproduction, the proton in a diffractive interaction is scattered at very small
angles and practically does not take part in the effective reaction. Analogously to
the emission of photons and to the idea of equivalent photon flux defined in QED,
one can think of hadron diffraction in terms of Pomeron emission and the ``Pomeron 
flux factor". This picture (and the IS model as a realization of it) has been 
successfully employed to
the hadron vertex in some HERA diffractive processes, such as leading baryon
production and diffractive DIS \cite{batista}, photoproduction \cite{foto},
and electroproduction \cite{eletro}.

However, such an approach is affected by a problem which is mostly concerned with 
its energy dependence. As it is
theoretically well known from very long, the Triple Pomeron integrated cross
section violates unitarity \cite{pdbcollins},  although its $x_{{\tt I\! P}}$
and $t$ dependences seem to be in good agreement with the available data 
\cite{dino2}.
In order to overcome this unitarity violation issue, we follow here
the ``renormalization" procedure originally
proposed in \cite{dino} and further discussed in \cite{dino2}, that is
\begin{eqnarray}
f_{\pom}(x_{\pom},t)=\frac{f(x_{\pom},t)}{\int_{x_{\pom_{min}}}^{x_{\pom_{max}}}
\int_{t=0}^{\infty}f(x_{\pom},t)\ dx_{\pom}\ dt}.
\label{erenorm}
\end{eqnarray}

For the ``unnormalized" flux factor $f(x_{\pom},t)$, we take the
Donnachie-Landshoff parametrization \cite{donna},
\begin{equation}
f(x_{\tt I\! P},t)=\frac{9{\beta}_{0}^{2}}{4{\pi}^2}
F_{1}^2(t)\ {x_{\tt I\! P}}^{1-2{\alpha}_{\pom}(t)}
\label{dlf}
\end{equation}
where $F_1(t)$ is the Dirac form factor,
\begin{equation}
F_1(t) = \frac{(4m^2-2.79t)}{(4m^2-t)}\ \frac{1}{(1-\frac{t}{0.71})^2}.
\end{equation}

Notice that, by choosing the renormalization procedure, ${\beta}_{0}$ does need
to be specified since it is crossed out as well as the other constant factors
appearing in Eq.~(\ref{dlf}). Yet about this equation, our choice for the Pomeron
trajectory has been $\alpha_{\pom}(t)=1.2 + 0.25\ t$, which is compatible with
both Tevatron and HERA data.

\subsection{The Pomeron Structure Function}

The Pomeron structure function has been established as a three-flavor quark
singlet at the initial scale, chosen to be $Q_0^2 = 2\ GeV^2$, with the gluon
component being generated by DGLAP evolution. Thus, no initial gluon distribution
has been assumed. The parametrization
used for the initial quark distribuition was
\begin{eqnarray}
\nonumber
\beta \Sigma(\beta,Q^2_0) = && [A_1 \exp(-A_2 \beta^2)+
B_1(1-\beta)^{B_2}]\ \beta^{0.001}\\
&& +\  C_1 \exp[-C_2(1- \beta)^2] (1-\beta)^{0.001},
\end{eqnarray}
which includes different amounts of soft, hard, and superhard profiles according to
the chosen parameters. The results presented below were obtained with the
following parameters: $A_1$ = 4.75 and $A_2$ = 228.4 for the soft part,
$B_1$ = 1.14 and $B_2$ = 0.55 for the hard one, and finally $C_1$ = 2.87 and
$C_2$ = 100 for the superhard term.

Wherever necessary, DGLAP evolution of the Pomeron parton densities has been
processed by using the program QCDNUM \cite{qcdnum}. For the proton
(or antiproton, when was the case), the parton densities were taken from
the parametrizations given in Ref.~\cite{gluck}.

%
%%%%%%%%%%%%%%%%%%%%%%%%%%%%%%%%%%%%%%%%%%%%%%%%%%%%%%%%%%%%%%%%%%%%%%%%%%
%

\section{Results and Discussion}

In the following, we present our predictions for hard diffractive production of
W's and dijets based on the previous discussion. These predictions are compared
with experimental data from Refs.~\cite{dzero,wcdf,rapgap} in Tables I-III.

%%%%%%%%%%%%%%%%%% TABLE 1
\begin{table}
\caption{\label{tab:table1} Data versus model predictions.
Diffractive W's and dijets were measured at $\sqrt{s} =$ 1800 GeV by the CDF
Collaboration \cite{wcdf,rapgap}. In both cases, $x_{\pom} < 0.1$ and
$E_{T_{min}} =$ 20 GeV. For the case of W production, $E_{T_{min}}$ refers
to the detected lepton while for dijet production it refers to the detected jet.}
\begin{ruledtabular}
\begin{tabular}{lccr}
Yield  & Rapidity  & Data (\%) & Model\\
\hline
W &    $-1.1<\eta_e<1.1$  & $1.15\pm 0.55$  & 0.35\\
jj &    $-3.5<\eta_j<-1.8$  & $0.75\pm 0.10$  & 0.72\\
\end{tabular}
\end{ruledtabular}
\end{table}
%%%%%%%%%%%%%%%%%%

%%%%%%%%%%%%%%%%% TABLE 2
\begin{table}
\caption{\label{table2} Data versus model results corresponding to the D0
experiment. The experimental data are from Ref.~\cite{dzero}
and the model calculations were performed with $E_{T_{min}} =$ 15 GeV for
$\sqrt{s} =$ 1800 GeV
and $E_{T_{min}} =$ 12 GeV for $\sqrt{s} =$ 630 GeV. In both cases,
$x_{\pom} < 0.1$. }
\begin{ruledtabular}
\begin{tabular}{lcccr}
$\sqrt{s}$ (GeV) & Rapidity  & Data (\%) &  Model\\
\hline
1800 & $|\eta|>1.6$  & $0.65\pm 0.04$  & 0.90\\
1800 & $|\eta|<1.0$  & $0.22\pm 0.05$  & 0.37\\
$\ 630$  & $|\eta|>1.6$  & $1.19\pm 0.08$  & 1.80\\
$\ 630$  & $|\eta|<1.0$  & $0.90\pm 0.06$  & 0.98\\
\end{tabular}
\end{ruledtabular}
\end{table}
%%%%%%%%%%%%%%%%%%%%%%%%

%%%%%%%%%%%%%%%% TABLE 3
\begin{table}
\caption{\label{table3} Experimental ratios versus model results corresponding
to the D0 experiment. Data are from Ref.~\cite{dzero}.}
\begin{ruledtabular}
\begin{tabular}{lcr}
 Ratios  & Data (\%) &  Model\\
\hline
630/1800    $|\eta|>1.6$  & $1.8\pm 0.2$  & 2.0\\
630/1800   $|\eta|<1.0$  & $4.1\pm 0.9$  & 2.7\\
1800 GeV     $|\eta|>1.6/|\eta|<1.0$  & $3.0\pm 0.7$  & 2.4\\
630 GeV   $|\eta|>1.6/|\eta|<1.0$  & $1.3\pm 0.1$  & 1.8\\
\end{tabular}
\end{ruledtabular}
\end{table}
%%%%%%%%%%%%%%%%%

In Table~I, the difficulty in obtaining a perfect and simultaneous description of 
both W and dijet production is evident. The situation is much better, however,
when one considers only jets. Besides the agreement
exhibited in Table I for the CDF experiment, consistency is also found with the
D0 results (Tables II and III).

In Table II both forward and central dijet production at two energies are
considered. For all cases, one sees that the model predictions are close to
the data, but slightly above. This sort of discrepancy is expected since effects
of experimental acceptance were not taken into account in these
predicitions. Such effects would certainly reduce these theoretically
predicted rates, but it is difficult to estimate to what extent.

In Table III is where the agreement between theory and data is generally better.
In this case, two kind of ratios are calculated: ratios between rates at different
energies but at the same rapidity range and the reverse, ratios between rates
taken at the same energy but different rapidity ranges. The better agreement
here could be attributed to the fact that these ratios would cancel the
normalization and acceptance effects to some extent.

In summary, we have shown that it is possible to obtain a reasonable
overall description of hard diffractive hadroproduction by a model based on the
Ingelman-Schlein approach once a quark-rich Pomeron structure function is assumed
and its DGLAP evolution performed. This result, {\it i.e.} the predominance of
quarks in the Pomeron ``valence" distribution, already obtained in \cite{mara},
is in conflict with the parametrizations independently established from HERA data
\cite{batista,foto,eletro}. This discrepancy may be seen as an additional
indication of factorization breaking \cite{collins} in hadronic diffraction.
However, if that is the real reason, it is quite intriguing that the consistence
between the data and theory shown here is possible at all.

%%%%%%%%%%%%%%%%%%%%%%%%%%%%%%%%%%%%%%%%%%%%%%%%%%%%%%%%%%%%%%%%%%%%%%%%%%%%

\section*{Acknowledgments}

We would like to thank the Brazilian governmental agencies CNPq and FAPESP
for their financial support.

%%%%%%%%%%%%%%%%%%%%%%%%%%%%%%%%%%%%%%%%%%%%%%%%%%%%%%%%%%%%%%%%%%%%%%%%

%
%%%%%%%%%%%%%%%%%%%%%%%%%%%%%%%%%%%%%%%%%%%%%%%%%%%%%%%
%

%%%%%%%%%%%%%%%%%%%%%%%%%%%%%%%%%%%%%%%%%%%%%%%%%%%%%%%%%%%%%%%%%%%%%%%%%%%%


\begin{references}
\bibitem{nota1} {We have presented a more detailed discussion about
this cross section in Ref.\cite{mara}. In Eq.~(\ref{dsw}) we have corrected two
misprints appearing therein.}
\end{references}
\end{document}